\documentclass[aps,prd,superscriptaddress,onecolumn,nopacs,nofootinbib,amsmath,amssymb,amsfonts]{revtex4}
\usepackage{dcolumn}      
\usepackage{bm}           

\usepackage{amsmath,amssymb}  
\usepackage{bm}  

\usepackage[pdftex,bookmarks,colorlinks,breaklinks]{hyperref}  

\usepackage{graphicx}
\usepackage{mathrsfs}
\usepackage{amsmath}

\newcommand{\be}{\begin{equation}}
\newcommand{\ee}{\end{equation}}
\newcommand{\ba}{\begin{eqnarray}}
\newcommand{\ea}{\end{eqnarray}}
\newcommand{\bs}{\begin{subequations}}
\newcommand{\es}{\end{subequations}}

\newcommand{\sfrac}[2]{{\textstyle\frac{#1}{#2}}}
\renewcommand{\:}[2]{{\textstyle\frac{#1}{#2}}}
\renewcommand{\;}[2]{{\frac{#1}{#2}}}

\newcommand{\forget}[1]{\iffalse#1\fi}
\newcommand{\forgetmenot}[1]{\iftrue#1\fi}

\newcommand{\del}{\nabla}
\renewcommand{\d}{\mathrm{d}}
\renewcommand{\div}{\hskip0.9pt{\mathsf{div}\hskip2pt}}
\newcommand{\Div}{\hskip0.9pt{\mathrm{div}\hskip2pt}}
\newcommand{\dv}{\Div}

\newcommand{\curl}{\hskip0.9pt{\mathsf{curl}\hskip2pt}}
\newcommand{\Curl}{\hskip0.9pt{\mathrm{curl}\hskip2pt}}
\newcommand{\dis}{\hskip0.9pt{\mathsf{dis}\hskip2pt}}
\newcommand{\Dis}{\hskip0.9pt{\mathrm{dis}\hskip2pt}}
\newcommand{\Del}{\hskip0.9pt{\mathrm{ D}\hskip0.5pt}}
\newcommand{\sdel}{\vec{\nabla}}
\newcommand{\Qs}{Q^{(S)}}
\newcommand{\Qv}{Q^{(V)}}
\newcommand{\Qvb}{{\bar Q}^{(V)}}
\newcommand{\Qt}{Q^{(T)}}
\newcommand{\Qtb}{{\bar Q}^{(T)}}

\renewcommand{\>}{\rangle}

\newcommand{\V}[1]{{\bm{#1}}}
\newcommand{\T}[1]{{\bm{#1}}}

\begin{document}

\title{Locally extracting scalar, vector and tensor modes in cosmological perturbation theory}

\author{Chris~Clarkson}
\email{chris.clarkson@uct.ac.za}
\affiliation{Astrophysics, Cosmology \& Gravitation Centre, and, Department of Mathematics and Applied Mathematics, University of Cape Town, Rondebosch 7701, Cape Town, South Africa}
\author{Bob Osano}
\email{bob.osano@uct.ac.za} 
\affiliation{Astrophysics, Cosmology \& Gravitation Centre, and, Department of Mathematics and Applied Mathematics, University of Cape Town, Rondebosch 7701, Cape Town, South Africa}

\begin{abstract}

Cosmological perturbation theory relies on the decomposition of perturbations into so-called scalar, vector and tensor modes. This decomposition is non-local and depends on unknowable boundary conditions. The non-locality is particularly important at second- and higher-order because perturbative modes are sourced by products of lower-oder modes, which must be integrated over all space in order to isolate each mode. However, given a trace-free rank-2 tensor, a locally defined scalar mode may be trivially derived by taking two divergences, which knocks out the vector and tensor degrees of freedom.  A similar local differential operation will return a pure vector mode. This means that scalar and vector degrees of freedom have local descriptions. The corresponding local extraction of the tensor mode is unknown however. We give it here. The operators we define are useful for defining gauge-invariant quantities at second-order. We perform much of our analysis using an index-free `vector-calculus' approach which makes manipulating tensor equations considerably simpler.

\end{abstract}

\date{\today}
\maketitle

\section{Introduction}

Perturbation theory in cosmology rests upon the decomposition of
perturbations into scalar, vector and tensor parts, on a background which has constant curvature. This is a generalization of Helmholtz's theorem to tensors on 3-spaces of constant curvature~\cite{Stewart:1974uz,Stewart:1990fm,Kodama:1985bj}. 
This split is usually performed non-locally: either in harmonic space or using an implicit integral over a Green's function (they are equivalent). Because of this, we must specify boundary conditions in order to define these non-local variables. Furthermore,  we must know a perturbation variable \emph{everywhere} in order to specify a given scalar vector or tensor type of perturbation \emph{somewhere}. For example, let us consider Helmhotz's theorem in 3-d flat space. Any vector $\bm V$ can be trivially written in terms of a scalar $\psi$ and vector $\bm A$:
\be
\bm V=\sdel \psi+\curl\bm A, 
\ee 
where 
\ba
\sdel^2\psi&=&\div\bm V,\\
\sdel^2\bm A&=&-\curl\bm V,
\ea
which follow from the standard vector calculus identities $\div\curl\bm V=0$ and $\curl\curl\bm V=-\sdel^2\bm V+\sdel\div\bm V$ (in Euclidean space).
Unique solutions for $\psi$ and $\bm A$ exist provided they vanish sufficiently fast at infinity. That is, $\psi$ and $\bm A$ are inherently non-local, requiring knowledge of $\bm V$ everywhere just to be defined at a single point. On the other hand, we can think of $\phi=\div\bm V$ as a pure scalar degree of freedom which is defined locally wherever $\bm V$ is; similarly, $\curl\bm V$ is a pure vector degree of freedom. So, given $\bm V$ we can isolate locally defined, unique, scalar and vector degrees of freedom by differentiating it appropriately. This is useful in electromagnetism where, for example, it is common to write down decoupled wave equations for the scalar and vector potentials which are equivalent to Maxwell's equations. Of course, the solution to these equations involves the same reliance on boundary conditions as defining $\psi$ and $\bm A$ do; yet the split of the dynamical wave equations themselves into decoupled solenoidal and irrotational parts is conceptually useful~-- and this doesn't have to rely on non-local conditions, or integrals over all space. 

Given the unique nature of cosmology, it seems perverse for us to only be able to define  gravitational waves or the gravitational potential utilising unknowable boundary conditions and conditions outside our horizon. The universe could be spatially infinite, implying that we cannot arbitrarily assign boundary conditions at infinity (which could depend on the start of inflation, for example). 
For scalar modes, however, we can reformulate perturbation theory into an equivalent local theory because we can simply take divergences to form objects which are locally defined pure scalars. Because of this they have a meaning which is well defined physically. Similarly for vectors. But what is the equivalent operation for tensor modes? Given a rank-2 tensor, how do we \emph{locally} isolate the tensor degrees of freedom? That is the goal of this paper. For readers who just can't wait to find out, the answer is given by Eq.~(\ref{tensor-ex}).

\section*{Notation and identities}

We assume an FLRW geometry of curvature $K$, and all relations are
defined on this background. Under perturbation at order $n$, all
relations below hold for objects of perturbative order $n$; for
lower perturbative order, the commutation relations below have
curvature correction terms added to them. Furthermore, we assume the objects we deal with are tensors in the full spacetime, which implies that they must be gauge-invariant objects at order $n$ from the Stewart-Walker lemma~\cite{Stewart:1974uz}. 

Given the usual
4-velocity $u^a$ we define the spatial metric
$h_{ab}=g_{ab}+u_au_b$ and volume element $\varepsilon_{abc}=u^d\eta_{abcd}$~\cite{Ellis:1998ct}. All other rank-1 and -2 tensors used here are
orthogonal to $u^a$, and rank-2 tensors are projected symmetric and
trace-free (PSTF) which we denote using angle brackets on indices.\footnote{Because of the confusing lexicon of scalar, vector and tensor modes, we will try to call things like a `vector' (a spatially projected 4-vector, which in an FLRW background is a 3-vector) a rank-1 tensor, and a `tensor' a `rank-2 tensor', etc. Ridiculous, but it should avoid confusion.}
We define the conformal (comoving) spatial covariant derivative acting on scalars or spatial
tensors as $\sdel_a=ah_a^{~b}\del_b=a\Del_a$, where $a$ is the scale
factor and $\Del_a$ is the spatial derivative normally used in the
covariant approach. \footnote{Strictly speaking $\Del_aX_{b\dots c}=h_a^{~a'}h_b^{~b'}\cdots h_c^{~c'}\del_{a'}X_{b'\dots c'}$, which we must use for objects which are not at maximal perturbative order. } We use $\sdel_a$ as it commutes with
$u^a\del_a$ and is the covariant derivative normally used in the
metric approach to perturbation theory when its index is downstairs (when $K=0$ it is just the partial derivative); all expressions below are covariant, and indices are raised and lowered with $g_{ab}$ (or $h_{ab}$ for PSTF objects). The irreducible parts of
the spatial derivative of PSTF tensors are the divergence, curl,
and distortion, defined as~\cite{Maartens:1996ch}
\ba
\div X_{b\dots c}&=& \sdel^a X_{ab\dots c}\\
\curl X_{ab\dots c}&=&\varepsilon_{de\<a}\sdel^d X_{b\dots c\>}^{~~~~~e}\\
\dis X_{ca\dots b}&=&\sdel_{\<c}X_{a\dots b\>}.
\ea
Then, the spatial derivative of a rank-$n$ PSTF tensor $X_{A_n}=X_{a_1a_2\ldots a_n}$ may be decomposed as (for $n=1,2,3$)\footnote{All our derivatives may be generalised to projected covariant derivatives in a general spacetime by replacing $\sdel\mapsto\Del$. Then we have the invariant parts of derivatives of PSTF tensors $\dv,\Curl~\&~\Dis$ which we discuss in the appendix. These don't commute with the derivative along $u^a$, even in an FLRW geometry. }
\be
\sdel_b X_{A_n}=\frac{2n-1}{2n+1}\,\div X_{\<A_{n-1}}h_{a_n\>b}
-\frac{n}{n+1}\,\curl X_{c\<A_{n-1}}\varepsilon_{a_n\>b}^{~~~~~c}
+\dis X_{bA_n}\,.
\ee 
Note that the divergence decreases the rank of the tensor by one,
the curl preserves it, while the distortion increases it by one (and all are PSTF).
Keeping this in mind one can drop the indices on differential
operators as long as it's explicit the valance of the PSTF tensor
which is being acted on (though this can sometimes be confusing). This
considerably simplifies the appearance of the equations.

There are important commutation relations for each of the three invariant derivatives which we need, which are derived from the 3-Ricci identity. The Riemann tensor on a 3-space of constant curvature is given by:
\be
{}^3\!R^{ab}_{~~\,cd}=\frac{1}{3} {}^3\!R\,\,h^a_{\,~[c}h^b_{\,~d]}=-\frac{2K}{a^2} \,h^a_{\,~[c}h^b_{\,~d]}~~~\Rightarrow~~~\sdel_{[a}\sdel_{b]}X_c={K}X_{[a}h_{b]c}\,
\ee
Then, when acting on a rank-1 spatial tensor (a 3-vector) the following relations hold:
\ba
\label{V1}\div\curl&=&0\label{divcurl-v}\\
\label{V2}\dis\curl&=&2\curl\dis\label{discurl-v}\\
\label{V3}\curl^2+\sdel^2&=&\sdel\div +2K\label{lap-curl-v}\\
\label{V4}\div\dis&=&\:12\sdel^2+\:16\sdel\div+K\label{divdis-v}\\
\label{V5}\sdel^2\div&=&\div\sdel^2-2K\div\\
\label{V6}\sdel^2\curl&=&\curl\sdel^2\label{lapcurl-v}\\
\label{V7}\sdel^2\dis&=&\dis\sdel^2+4K\dis\label{lapdis-v}
\ea
Here we have used the notation whereby $\sdel$ with no index represents the gradient of a scalar. These are the familiar vector calculus identities, on spaces of constant curvature. To be clear about notation, the first of these is $\varepsilon_{abc}\sdel^a\sdel^bX^c=0$, and the second $\sdel_{\<a}(\varepsilon_{b\>cd}\sdel^cX^d)=\varepsilon_{de\<a}\sdel^d\sdel_{b\>}X^e+\varepsilon_{de\<a}\sdel^d\sdel^eX_{b\>}-\frac{2}{3}\varepsilon_{d\<ab\>}\sdel^d\sdel^fX_f=\varepsilon_{de\<a}\sdel^d\sdel_{b\>}X^e$. Note also that the first is a rank-0 tensor equation (curl preserves rank, div reduces by one), while the second is rank-2 (dis increases by one).

Acting on any PSTF rank-2 tensor, the following commutation
relations hold:
\ba
\label{T2}\curl\div&=&2\div\curl\label{curldiv-t}\\
\label{T3}\curl^2+\sdel^2&=&\:32\dis\div +3K\label{lap-curl-t}\\
\label{T9}\dis\div&=&\:52\div\dis-\:56\sdel^2-5K\label{disdiv-t}\\
\label{T4}(\dis\div)\curl&=&\curl(\dis\div)\\
\label{T8}\div\curl\div&=&0 \label{divcurldiv}\\
\label{T1}\sdel^2\curl&=&\curl\sdel^2\\
\label{T5}\sdel^2\div&=&\div\sdel^2-4K\div\label{lapdiv-t}\\
\label{T6}\sdel^2\div\div&=&\div\div\sdel^2-6K\div\div
\ea
where Eq.~(\ref{T9}) is given in~\cite{Maartens:1996ch}. Some of these are really tedious to derive, but are easy to use in this form. Note that both $\sdel^2$ and $\curl^2$ can be written in terms of $\dis\div$ and $\div\dis$. 

\section{The decomposition theorems}

We discuss now the decomposition of rank-1 and -2 (PSTF) tensors into their SVT parts. In particular, we shall separate local SVT variables from non-local ones. We shall denote local SVT variables by a breve $\breve{~}$. Our notation will be such that the same letter with a different number of indices represent different but usually related objects (e.g., $S_a=\sdel_a S$ and $S_{ab}=\sdel_{\<a} S_{b\>}=\sdel_{\<a} \sdel_{b\>}S$ etc.).

\subsection{rank-1 tensors}

For rank-1 tensors the scalar and vector parts correspond to the
curl- and divergence-free parts respectively. Given a spatial
rank-1 tensor on an FLRW background at maximum perturbative order,
$X^a=S^a+V^a=\sdel^aS+V^a$, it is easy to form rank-1 tensor quantities which
are pure scalars or vectors by the following rules:
\ba
\text{Scalar:}~~~&&  \breve S_a\equiv\sdel_a\div X = \sdel_a\div S = \sdel_a\sdel^2 S \nonumber\\
&&~~~~~~=    \left(\sdel^2-2K\right)\sdel_a S\nonumber\\
\text{Vector:}~~~&&  \breve V_a\equiv\curl X_a=\curl
V_a.\label{sv}
\ea
By taking spatial derivatives we have quantities which are scalars
and vectors but which remain local. We can then formulate the
non-local variables from the original tensor through these variables by the formal
solution
\ba
S&=&\sdel^{-2}\div X=\sdel^{-2}\sdel^{-2}\div\breve S\,,\\
V_a&=&-\left(\sdel^2-2K \right)^{-1}\curl^{2} X_a=-\left(\sdel^2-2K \right)^{-1}\curl \breve V_a\,.\label{vecvec}
\ea
Here we have used standard inverse Laplacian notation whereby $\left(\sdel^2-nK \right)^{-1}$ stands for the solution of the corresponding elliptic differential equation: assumptions about behaviour at infinity or boundary conditions must be made. While $X^a$, $\breve S^a$ and $\breve
V^a$ have compact support, $S$ and hence $V^a$ do not~\cite{Stewart:1990fm}. 

Part of
the conceptual utility of defining local scalar and vector
quantities via Eqs.~(\ref{sv}), which are the same rank as $X^a$,
is that the differential operations involved commute with the
Laplacian $\sdel^2$, and time derivative. Therefore, if $X^a$
satisfies a wave equation with source,
$\mathcal{L}[X_a]=\mathcal{S}_a$, where $\mathcal{L}$ is a linear
differential operator containing the Laplacian, then we may
locally extract the covariant scalar and vector parts to find
$\mathcal{L}[\breve S_a]=\sdel_a\div\mathcal{S}$ and
$\mathcal{L}[\breve V_a]=\curl
\mathcal{S}_a$.

One may further relate the local and non-local decompositions in
Fourier space (which is inherently non-local, and relies on the functions having compact support~-- i.e., vanishing sufficiently fast at infinity). Defining a scalar
harmonic basis in the usual way, $\sdel^2\Qs=-k^2\Qs$, we find
$\breve S^{(k)}=-k^2 S^{(k)}$. Similarly for vectors,
$\sdel^2\Qv_a=-k^2\Qv_a$, where we have two parities of orthogonal
harmonics, $(k^2+2K)^{1/2}\Qv=\curl\Qvb\Leftrightarrow
(k^2+2K)^{1/2}\Qvb=\curl\Qv$, in Fourier space the local
extraction involves a parity switch. However, this feature can be
trivially removed by defining $\breve V_a\equiv\curl\curl X_a$
instead of Eq.~(\ref{vecvec}).

\subsection{rank-2 tensors}

For rank-1 tensors the above considerations are trivial to find
and well known. For rank-2 tensors, on the other hand, the local
decomposition into scalar, vector and tensor modes is not quite so
easy. A general projected, symmetric, and trace-free rank-2 tensor
has the decomposition
\ba
X_{ab}&=&S_{ab}+V_{ab}+T_{ab}\nonumber\\
&=&\sdel_{\<a}\sdel_{b\>}S+\sdel_{\<a}V_{b\>}+T_{ab},
\ea
where the non-local scalar part $S_{ab}$ is curl-free, the vector part $V_{ab}$ is
solenoidal, $\div V=0\Rightarrow\div\div V=0$ (and $\div
V_a\neq0$~-- note the notation: $\sdel^aV_a=0\Rightarrow\sdel^a\sdel^bV_{ab}=\sdel^a\sdel^b\sdel_{\<a}V_{b\>}=0$), while the tensor part is transverse, $\div T_{a}=0$.
The question is, how do we form \emph{local} scalar, vector and
tensor quantities from $X_{ab}$ and relate them to the non-local
split given above? That is, what differential operations do we need to do $X_{ab}$ to leave only the scalar, vector or tensor part?

Furthermore, let us assume that $X_{ab}$ obeys
a wave equation with source of the form
\be
\mathcal{L}[X_{ab}]=\mathcal{S}_{ab}\label{wave}
\ee
where $\cal L$ contains time derivatives and Laplacians, and any derivative operations which preserve the rank of $X_{ab}$~-- i.e., $\curl$, $\dis\div$ or $\div\dis$, or combinations thereof.
Ideally we would like the local extractions to be differential
operators which commute with $\cal L$. To show this we need only show that the Laplacian and $\curl$ commute with our extractions below: time derivatives commute trivially; $\dis\div$ commutes if $\curl$ and $\sdel^{2}$ does by Eq.~(\ref{lap-curl-t}); and $\div\dis$ therefore will by Eq.~(\ref{disdiv-t}).

\subsection{scalars}

 Clearly, $\breve S=\div\div X$ is a
(covariantly defined) scalar and can only depend on $S$, so let us
define
\ba
\breve S_{ab}&\equiv&\hat{\mathscr{S}}[X_{ab}]\equiv\sdel_{\<a}\sdel_{b\>}\div\div
X=\sdel_{\<a}\sdel_{b\>}\sdel^c\sdel^d X_{cd}
\nonumber\\
&=&\sdel_{\<a}\sdel_{b\>}\div\div
S\nonumber\\
&=&\:23\sdel_{\<a}\sdel_{b\>}\left(\sdel^2+3K\right)\sdel^2S.
\ea
The non-local scalar $S$ is given formally from $X_{ab}$ by
\ba
\label{ScaExt1}S&=&\;32\left(\sdel^2+3K\right)^{-1}\sdel^{-2}\div\div X\nonumber\\
&=&\;32\left(\sdel^2+3K\right)^{-1}\sdel^{-2}\breve S.
\ea
This latter relation trivially gives the relation in Fourier space
by replacing $\sdel^2\mapsto -k^2$.

Defining $\breve S_{ab}$ to preserve the rank of the original tensor allows us to find the wave equation it satisfies easily. First, note that any curls in $\mathcal{L}$ commute with $\hat{\mathscr{S}}$ trivially (producing zero), and note that (non-trivially)
\be
\sdel^2\left(\sdel_{\<a}\sdel_{b\>}\div\div\right)=\left(\sdel_{\<a}\sdel_{b\>}\div\div\right)\sdel^2.
\ee
We then see that the locally defined scalar quantity
$\breve S_{ab}$ obeys
\be
\mathcal{L}[\breve S_{ab}]=\hat{\mathscr{S}}[\mathcal{S}_{ab}]=\sdel_{\<a}\sdel_{b\>}\div\div \mathcal
S.
\ee

\subsection{vectors}\label{vectors}

We begin by noting that the operator
$\curl\div=2\div\curl$ knocks out the scalar and tensor part of
$X_{ab}$, and is solenoidal by Eq.~(\ref{divcurldiv}) (note that
it forms a rank-1 tensor from $X_{ab}$); thus, $\breve
V_a=\curl\div X_a$ is a locally defined vector. The rank-2
extraction of vector modes may be given by taking the distortion
of this operator,
\ba
\label{vecExt1}\breve V_{ab}&\equiv&\hat{\mathscr{V}}[X_{ab}]\equiv \dis\curl\div X_{ab}\nonumber\\
&=&\varepsilon_{cd\<a}\sdel_{b\>}\sdel^c\sdel^e
X_e^{~d}\nonumber\\
&=&\dis\curl\div\dis
V_{ab}\\
&=&\:12\dis\curl\left(\sdel^2+2K\right) V_{ab}.
\ea
The non-local vector $V_a$ may be given formally from $X_{ab}$ by
\ba
V_a&=&2\left(-\sdel^4+4K^2\right)^{-1}\curl^2\div X_a\nonumber\\
&=&2\left(-\sdel^4+4K^2\right)^{-1}\curl\breve V_a.
\ea
This last relation enables us to relate the local and non-local
vectors in Fourier space as $\breve
V^{(k)}=\:12(k^2+2K)^{1/2}(-k^2+2K)\bar{V}^{(k)}$, with the same
relation for the opposing parity.

Again, the otherwise superfluous preservation of the rank of
$X_{ab}$ in defining the operator $\hat{\mathscr{V}}$~-- the extra
$\dis$ term~-- allows $\hat{\mathscr{V}}$ and $\mathcal{L}$ to
commute [to prove this for the Laplacian, we use Eqs. (\ref{lapdis-v}),
(\ref{lapcurl-v}) and (\ref{lapdiv-t}); for curl, we use Eqs.~(\ref{curldiv-t}) followed by ~(\ref{discurl-v})], giving our wave equation
for $\breve V_{ab}$ as
\be
\mathcal{L}[\breve
V_{ab}]=\hat{\mathscr{V}}[\mathcal{S}_{ab}]=\dis\curl\div\mathcal{S}_{ab}.
\ee

\subsection{tensors}

The key difficulty here lies in finding the differential
operator which when acting on $\dis V_{ab}=\sdel_{\<a}V_{b\>}$ produces zero by
virtue of $V_a$ being solenoidal (while obviously leaving the
transverse part of $X_{ab}$). It is straightforward to verify that
$[\curl^2+\:12\dis\div-K]\dis V_{ab}=0$ for $\div V=0$, which
follows from Eqs. (\ref{discurl-v}),  (\ref{divdis-v}) and
(\ref{lap-curl-v}). We must also take an extra curl to remove the
scalar part. Therefore, our local tensor extraction may be defined
as
\ba
\breve
T_{ab}&\equiv&\hat{\mathscr{T}}[X_{ab}]\equiv\left[
-\sdel^2+2K+2\dis\div\right]\curl X_{ab}\nonumber\\
 &=&\varepsilon_{cd\<a}\left[\left( -\sdel^2+2K\right)\sdel^c
X_{b\>}^{~d}+\sdel_{b\>}\sdel^e\sdel^cX_{e}^{~d}\right]
\nonumber\\&&+\varepsilon_{cde}\sdel_{\<a}\sdel^e\sdel^cX_{b\>}^{~d}\nonumber\\
&=&\left[ -\sdel^2+2K\right]\curl T_{ab}.\label{tensor-ex}
\ea
Note that $\curl$ commutes with the operator in square brackets.
It is relatively straightforward to verify that $\breve T_{ab}$ is transverse,
$\div\breve T_a=0$, showing that this represents a tensor mode as expected: First note that we may write
\be
\hat{\mathscr{T}}=\:13\left[\sdel^2+4\curl^2-6K\right]\curl ,\label{tensor2}
\ee
 which follows from Eq.~(\ref{lap-curl-t}), so that
\ba
\label{divzero}\div\hat{\mathscr{T}}&=&\:13\left[\sdel^{2}+\curl^{2}-2K\right]\div\curl\nonumber\\
&=&\:13\sdel\div\div\curl=0,
\ea
which uses Eqs.~(\ref{curldiv-t}) and~(\ref{lapdiv-t}) on the first line, then Eq.~(\ref{lap-curl-v}) to get to the second, and finally Eq.~(\ref{divcurl-v}) to show the last expression is zero.

The formal, non-local, TT tensor $T_{ab}$ is given in terms of the original tensor $X_{ab}$ by taking a further $\curl$ to obtain
\be
T_{ab}=\left(-\sdel^{2}+3K\right)^{-1}\left(-\sdel^{2}+2K\right)^{-1}
\hat{\mathscr{T}}[\curl X_{ab}].
\ee
To convert our extraction into Fourier space we define tensor harmonics as
$\sdel^2\Qt_{ab}=-k^2\Qt_{ab}$, where we have two parities of orthogonal
harmonics, $(k^2+3K)^{1/2}\Qt_{ab}=\curl\Qtb_{ab}\Leftrightarrow
(k^2+3K)^{1/2}\Qtb_{ab}=\curl\Qt_{ab}$~\cite{Tsagas:2007yx}. We therefore find
\ba
\breve T^{(k)}=\left(k^{2}+3K\right)^{1/2}\left(k^{2}+2K\right)\bar{T}^{(k)},
\ea
with the same relation for the opposite parity.

In order for $\hat{\mathscr{T}}$ to be useful it will have to operate on a wave equation, Eq.~(\ref{wave}). It is clear from the form of $\hat{\mathscr{T}}$ given in Eq.~(\ref{tensor2}) that $\mathcal{L}$ and $\hat{\mathscr{T}}$ commute, so that $\breve T_{ab}$ obeys
\be
\label{OP}\mathcal{L}[\breve T_{ab}]=\hat{\mathscr{T}}[\mathcal{S}_{ab}]=\left[
-\sdel^2+2K+2\dis\div\right]\curl \mathcal{S}_{ab}.
\ee

\section{Vector and tensor modes induced by scalars} 

A useful application of our extraction operators may be illustrated by considering second-order modes in cosmology which are induced by first-order scalar modes. We shall investigate this both in the covariant approach, and briefly in the metric based approach. In the former we shall see how the local extraction operators can help in the construction of gauge-invariant objects at second-order.

\subsection{The covariant approach}

We shall now consider how to use the relations we have derived in practice, using the 1+3 covariant formulation of perturbation theory for illustration~\cite{Ellis:1998ct,Tsagas:2007yx}. The covariant approach is adapted for perturbation theory because it deals with physically defined tensorial objects. Because of this, at a given perturbative order, any tensor which vanishes at lower order is automatically gauge-invariant by an extension of the Stewart-Walker Lemma~\cite{Stewart:1974uz,Bruni:1996im} (though they are not are not frame-invariant as an observer must be specified). This helps us to use the local SVT operators to form gauge-invariant quantities at second-order. 

We  introduce an index-free formulation of the covariant approach, as described in the Appendix. In this section the spatial derivative operators are the normal ones used in the covariant approach, and do not commute with the time derivative.

Consider a curved FLRW background with some fluid, with an equation of state $p=w\rho$. Now perturb this at first-order exciting scalar modes only. This can be described by the shear and electric Weyl curvature only:
\ba
\dot{\T\sigma}-\Dis\V A&=&-2H\T\sigma-\T{E}\,,\\
\dot{\T E}&=&-\:12(1+w)\rho\,\T\sigma-3H\T E\,,
\ea
where the acceleration is related to $\T E$ by
\be
\V A=-\frac{3w}{(1+w)\rho}\dv\T E\,.
\ee
The solution to this system determines all other gauge-invariant quantities, such as $\Del H=\:12\Div\T\sigma$, $\Del\rho=3\dv \T E$, etc. All these quantities are gauge-invariant by the Stewart-Walker Lemma. The fact that only scalars are allowed is actually signalled by the condition that $\Curl\T\sigma=0=\Curl\T E$, and so $\dv\Curl\T\sigma=0=\dv\Curl\T E$, which all follow when the vorticity is zero (at first-order).\footnote{When vectors and tensors are not set to zero we can use $\hat{\mathscr{S}}$, $\hat{\mathscr{V}}$ and
$\hat{\mathscr{T}}$ acting on $\T\sigma$ or $\T E$ to invariantly extract the SVT parts. }

If we now allow for second-order perturbations, then $\T\sigma$ and $\T E$ are no longer gauge-invariant, and contain a mixture of scalars, vectors and tensors, all sourced by the scalars at first-order (that is, the solutions to the first-order equations above). However, given our method of a local SVT extraction, we can form objects which are gauge-invariant at second-order: the rank-2 tensors
$\hat{\mathscr{V}}(\T\sigma)$, $\hat{\mathscr{V}}(\T E)$, $\hat{\mathscr{T}}(\T \sigma)$ and $\hat{\mathscr{T}}(\T E)$ are all gauge-invariant at second-order by the Stewart-Walker Lemma, since they vanish at first-order. It is important to note, however, that these objects may not necessarily  describe vectors and tensors because $\T\sigma$ and $\T E$ do not vanish at first-order, a key requirement in our proofs (e.g., $\hat{\mathscr{T}}(\T E)$ may not be divergence-free).\footnote{This is analogous to the situation at first-order whereby $\Del H$ or $\Del\rho$ can contain vector degrees of freedom when the vorticity is non-zero, despite appearing to be gradients of scalars. Since $H$ and $\rho$ are non-zero in the background, $\Del$ is not a pure gradient operator at first-order, and so $\Curl\Del\rho\propto\V\omega\neq0$. } A corollary to this is that because $\hat{\mathscr{V}}(\T\sigma)$, $\hat{\mathscr{V}}(\T E)$, $\hat{\mathscr{T}}(\T \sigma)$ and $\hat{\mathscr{T}}(\T E)$ form a set of second-order gauge invariant rank-2 tensors, we \emph{can} construct pure scalar, vector or tensor modes from them by operating further with $\hat{\mathscr{S}}$, $\hat{\mathscr{V}}$ or 
$\hat{\mathscr{T}}$ respectively. For example, $\hat{\mathscr{S}}[\hat{\mathscr{V}}(\T\sigma)]$ is a second-order gauge-invariant scalar degree of freedom, and $\hat{\mathscr{T}}[\hat{\mathscr{V}}(\T E)]$ is a second-order gauge-invariant tensor mode. These are quite cumbersome to use, however, because the objects they are formed from~-- $\T\sigma$ and $\T E$~-- are not gauge-invariant at second-order.

The magnetic Weyl curvature, on the other hand, is a fully gauge-invariant variable at second-order because it vanishes at first-order, and so is ideal for local SVT decomposition at second-order. We have 
\be
{\T{H}} =  -\Dis\V{\omega} +\Curl{\T{\sigma}}
\ee
which tells us that this contains both vectors (signalled by the vorticity) and tensor modes ($\Curl\T\sigma$ is zero for scalars at first-order, so gauge-invariantly signals tensor modes at second~\cite{Clarkson:2003af}).  Does $\T H$ contain scalars? No. Consider Eq.~(\ref{conR1}), 
\be
\dv {\T{H}}  = -\left(1+w\right)\rho\,\V{\omega}
- {\T{\sigma}}\times
{\T{E}}\,;
\ee
when we take the divergence of this, we find, using Eq.~(\ref{B28})
\be
\dv\dv {\T{H}}  = -\left(1+w\right)\rho\,\dv\V{\omega}
+ {\T{\sigma}}\cdot\Curl{\T{E}}-{\T{E}}\cdot\Curl{\T{\sigma}}
\ee
and this is identically zero at second-order. If we are to consider vector modes, then there are two sources: one is from the vorticity, but this isn't a generated vector mode at second-order, since it is straightforward to show that $\dot{\V\omega}=(3w-2)\,\V\omega$~\cite{Lu:2008ju}, on using $\Curl\V A=6wH\V\omega$. Let us set this to zero. The other vector degree of freedom may be found by locally extracting it from $\T{H}$; using the results of Sec.~\ref{vectors}, and Eq.~(\ref{curlxxy}) we have:
\be
\Curl\dv\T{H}=\:{7}{10}(\T\sigma\cdot\dv\T E-\T E\cdot\dv\T \sigma)+\T\sigma\cdot\Dis\T E-\T E\cdot\Dis\T \sigma\,.
\ee
Note that $\Curl\dv\T{H}$ is a rank-1 tensor describing a pure vector mode which is induced by first-order scalars, as we would hope. 

Now let us consider tensors, the tricky part. It is clear that $\T H$ can represent pure tensor modes only  when $\T\sigma\times \T E=0=\V{\omega}$. More generally, the tensor degrees of freedom are given by a wave equation sourced by scalars. Using the identity 
\ba
(\Curl\T{X}\dot)&=&\Curl\dot{\T{X}}-H\Curl\T{X}
-\sfrac{3}{10}\T{\sigma}\times\Div\T{X}
+\T\sigma\circ\Curl\T{X}
-\T\sigma\times\Dis\T{X}
+\T A\times\dot{\T X}+H\T A\times\T X
\ea
for $\T{E}$,
we find that $\T H$ satisfies the general wave equation at second-order (i.e., neglecting terms of order $\T H^2, \T\sigma\T H$, etc. and keeping $\V\omega=0$):
\ba
\label{wave1}
\ddot{\T{H}}-\Del^{2}\T{H} +7H\dot{\T{H}}+\left(6H^2-2w\rho+\frac{6K}{a^2}\right)\T{H} = {\bm{\mathcal{S}}},
\ea 
where the source $\bm{\mathcal{S}}$ is given by
\ba
\label{source1}
{\bm{\mathcal{S}}}=-\left(\:65+\:32w\right)\,\T{\sigma}\times\Div\T{E}-2\,\T{\sigma}\times\Dis\T{E}
+\:32w\T\sigma\times\div\T E
+3\left(-\:{4}{5}+w\right)\T{E}\times\Div\T{\sigma}+\frac{6w(2-3w)}{(1+w)}\frac{H}{\rho}\,\T{E}\times\Div\T{E}
\ea 
where the $\div\T{H}$ constraint, Eq.~(\ref{conR1}) (which allows a $\Dis\Div\T H$ term to be expressed in terms of first order product) is used simplify the first term in the source, which is subsequently expanded using identity Eq.~(\ref{B28}). 
Now, to get the gauge-invariant equation governing the tensor part of this equation it is a straightforward matter of applying the tensor extraction operator to this equation. Defining the tensor
\be
\T{\mathcal{H}}=a^{-3}\hat{\mathscr{T}}(\T{H})=\left[
-\Del^2+2 a^{-2}K+2\Dis\dv\right]\Curl \T{H} 
\ee
we have,
\ba
\label{wave2}
\ddot{\T{\mathcal{H}}}-\Del^{2}\T{\mathcal H} +13\,H\dot{\T{\mathcal H}}+\left[33H^2-\frac{1}{2}(1+7w)\rho+\Lambda+\frac{6K}{a^2}\right]\T{\mathcal H} = \left[
-\Del^2+\frac{2K}{ a}+2\Dis\dv\right]\Curl{\bm{\mathcal{S}}},
\ea 
which locally represents the gravitational wave part of the perturbations. The source terms can be further expanded into their invariant symmetric trace-free parts using the identities in the appendix.

\subsection{The metric approach}

To finish off, let us briefly consider how our approach can be applied in the metric approach to perturbations, in the same physical situation considered above where vectors and gravitational waves are sourced by scalar modes. This was considered in~\cite{Mollerach:2003nq,Ananda:2006af,Baumann:2007zm,Lu:2007cj,Lu:2008ju,Christopherson:2009bt}. In the Poisson gauge, using conformal time $\eta$,
\begin{equation}
ds^2 = -a^2 \left[ 1 + 2 \Phi\right] d\eta^2 - a^2
V_{i}dx^id\eta + a^2  \left [(1 - 2 \Phi)\gamma_{ij}+h_{ij} \right ]
dx^idx^j\,, \label{perturbed metric}
\end{equation}
where it is assumed that $\Phi$ is first-order (and there is no anisotropic stress) and $V_i$ and $h_{ij}$ are the second-order vector and tensor degrees of freedom in the metric. From the $ij$ part of the field equations we have
\ba
2\partial_{(i}V_{j)}'+ 4\mathcal{H}\partial_{(i}V_{j)}
+h''_{ij} + 2 \mathcal{H} h'_{ij} - \sdel^2 h_{ij}
=  {\cal S}_{ij} \label{ask}
\ea
where $\mathcal{H}=a'/a$, a prime denotes $\partial_\eta$, and the source is given, for an arbitrary constant equation of state $w$, by 
\ba
{\cal S}_{ij}
&=& -16 \Phi \partial_{\<i} \partial_{j\>} \Phi     -8\partial_{\<i} \Phi \partial_{j\>} \Phi 
     + \frac{16}{3(1+w) \mathcal{H}^2} \partial_{\<i} \left( \Phi' + \mathcal{H} \Phi \right)
      \partial_{j\>} \left( \Phi' + \mathcal{H} \Phi \right)\,.
\ea
The vector and tensor parts typically have to be separated by invoking an extraction operation which involves a non-local projection in Fourier space. For example, to extract the divergence free vector $V_i$ from
$\partial_{(i} V_{j)}$, we can use the operator
${\mathcal{V}}_{m}^{ij}$~\cite{Lu:2007cj},
\begin{equation}
{\mathcal{V}}_{m}^{ij}(\bm{x}, \bm{x}') = -\frac{2 i}{(2
\pi)^3} \int \d^3 k'\ {k'^{-2}} \int \d^3 x'\ {k'^{i}} \left [
e_{m}(\bm{k}') e^{j}(\bm{k}') + \bar{e}_{m}(\bm{k}')
\bar{e}^{j}(\bm{k}') \right ]  e^{i
\bm{k}'\cdot(\bm{x} - \bm{x}')},
\end{equation}
where $\bm e$ and $\bar{\bm e}$ are vectors orthogonal to $\bm k$ in Fourier space. 
Then $V_m(\bm{x}) = {\mathcal{V}}_{m}^{ij}(\bm{x},
\bm{x}') \partial_{(i} V_{j)} (\bm{x}') $. A similar operation isolates the tensor parts in a similar fashion: decompose the object in Fourier space, project out the  tensor degrees of freedom, and then reconstruct  a tensor object in real space. With such operators Eq.~(\ref{ask}) can be decomposed into two equations for each mode.

To perform the same operation locally is impossible. However, to use the formalism presented here is straightforward, and gives an alternative way to separate Eq.~(\ref{ask}) into a vector equation and a tensor one. Let us consider how to form the vector equation. Let us first convert Eq.~(\ref{ask}) into index-free notation:
\ba
 2\dis\V{V}'+4\mathcal{H}\dis\V{V}+\T{h}''+2\mathcal{H}\T{h}'-\del^2\T{h}&=&-16\Phi\dis\sdel\Phi-8\sdel\Phi\circ\sdel\Phi
\nonumber\\&& + \frac{16}{3(1+w) \mathcal{H}^2} \sdel\!\left( \Phi' + \mathcal{H} \Phi \right)\circ\sdel\!\left( \Phi' + \mathcal{H} \Phi \right)\,.
\ea
Now, if we operate with $\hat{\mathscr{V}}$ on both sides we have an evolution equation for $\hat{\mathscr{V}}(\dis\V{V})$:
\be
\left[a^2\hat{\mathscr{V}}(\dis\V{V})\right]'=a^2\hat{\mathscr{V}}\left[-8\Phi\dis\sdel\Phi-4\sdel\Phi\circ\sdel\Phi
+ \frac{8}{3(1+w) \mathcal{H}^2} \sdel\!\left( \Phi' + \mathcal{H} \Phi \right)\circ\sdel\!\left( \Phi' + \mathcal{H} \Phi \right)\right]\,.
\ee
The right hand side can be converted into irreducible form as follows. First note that $\hat{\mathscr{V}}(\Phi\dis\sdel\Phi)=-\hat{\mathscr{V}}(\sdel\Phi\circ\sdel\Phi)$, which implies that each term on the rhs is of the form $\hat{\mathscr{V}}(\sdel\Phi\circ\sdel\Psi)$. Using the identities in the appendix, together with the fact that $\curl\sdel\Phi=0$, we find that:
\ba
\dis\curl\div(\sdel\Phi\circ\sdel\Psi)&=&
\frac{5}{9}\left[\sdel\sdel^2\Phi\times\dis\sdel\Psi
+\sdel\sdel^2\Psi\times\dis\sdel\Phi
-\sdel\Phi\times\dis\sdel\sdel^2\Psi
-\sdel\Psi\times\dis\sdel\sdel^2\Phi
\right]
\nonumber\\
&+&\frac{1}{10}\left[\dis\sdel\Phi\times\div\dis\sdel\Psi
-\dis\sdel\Psi\times\div\dis\sdel\Phi
\right]
\nonumber\\
&-&\frac{1}{3}\left[\dis\sdel\Phi\times\dis\dis\sdel\Psi
-\dis\sdel\Psi\times\dis\dis\sdel\Phi
\right]\,.
\ea
Note that the last two lines vanish in the case when $\Psi=\Phi$. Hence, in the case of dust, or purely growing mode initial conditions where $\Phi'\propto\Phi$, these do not contribute to the induced vector mode. In these cases we have,
\ba
\hat{\mathscr{V}}(\dis\V{V})=\frac{40}{27(1+w)a^2}\int\d\eta\frac{a^2}{\mathcal{H}^2}\bigg\{
3(1+w)\mathcal{H}^2\left[\sdel\sdel^2\Phi\times\dis\sdel\Phi-\sdel\Phi\times\dis\sdel\sdel^2\Phi\right]\nonumber\\
+2\sdel\sdel^2(\Phi'+\mathcal{H}\Phi)\times\dis\sdel(\Phi'+\mathcal{H}\Phi)
-2\sdel(\Phi'+\mathcal{H}\Phi)\times\dis\sdel\sdel^2(\Phi'+\mathcal{H}\Phi)
\bigg\}\,.
\ea

A similar equation for the tensor part may be found in an analogous fashion.

\section{Discussion}

We have presented a novel way to invoke the scalar, vector and tensor split at the heart of cosmological perturbation theory in a local fashion without relying on boundary conditions. Although it appears calculationally cumbersome, this is conceptually important in cosmology because the boundary conditions are fundamentally unknowable. We have also developed a fully index-free `vector-calculus'-like approach for the 1+3 covariant approach which makes complicated tensor equations simpler to look at and much easier to manipulate. 

As an application we have examined the case of vectors and tensors which are induced at second-order from first-order density perturbations, and have seen that $\hat{\mathscr{V}}(\T H)$ and $\hat{\mathscr{T}}(\T H)$ are gauge-invariant variables which are a pure vector and tensor mode respectively. The covariant approach has a significant advantage over the metric approach in this regard: it is much harder to isolate gauge-invariant quantities at second-order starting from a non-covariant split of the metric, although we have discussed how to extract local SVT quantities using the same operations. While it is mathematically clear what is going on in this example, the physical interpretation of $\hat{\mathscr{V}}(\T H)$ and $\hat{\mathscr{T}}(\T H)$ is not obvious and deserves further investigation. It is worth noting that in Fourier space $\Curl \Div \T H$ and $\hat{\mathscr{T}}(\T H)$ are related to the usual non-local vector and tensor parts of $\T H$ by $k^3$ factors only, so the physical interpretation of the local SVT variables may not be much different from the usual ones. 

In this example, we have not considered the more involved problem of second-order scalars induced by first-order scalars. $\T H$ contains no scalar modes in this scenario. However, we have mentioned that we can define second-order gauge-invariant scalar modes by noting that $\hat{\mathscr{V}}(\T\sigma)$, $\hat{\mathscr{V}}(\T E)$, $\hat{\mathscr{T}}(\T \sigma)$ and $\hat{\mathscr{T}}(\T E)$ are rank-2 tensors at second-order which are gauge-invariant (since they all vanish at first-order), so these can be used as the basis to form  scalar modes at second-order (i.e., $\hat{\mathscr{S}}[\hat{\mathscr{V}}(\T\sigma)]$, $\hat{\mathscr{S}}[\hat{\mathscr{V}}(\T E)]$, $\hat{\mathscr{S}}[\hat{\mathscr{T}}(\T \sigma)]$ and $\hat{\mathscr{S}}[\hat{\mathscr{T}}(\T E)]$), which necessarily vanish at first-order, and are thus gauge-invariant. 

It is a more complicated issue to isolate pure second-order SVT modes locally when scalars, vectors and tensors are all present at first-order. At a given perturbative order, the local extraction operations only have the SVT interpretation when acting on a quantity that vanishes at lower orders (and is thus gauge-invariant at the relevant order if it is a covariantly defined object). One way to approach this is  to consider something like $\Dis\V\omega$: this cannot contain tensors at first-order since it is formed from a rank-1 tensor. Therefore, $\hat{\mathscr{T}}(\Dis\V\omega)$ must be a gauge-invariant object at second-order (provided it is non-zero). This is analogous to having vector modes at first-order contained in quantities such as $\Del\rho$ which look like scalars but are not in general. Similarly, quantities such as $\Dis\Del\Div\Div\T\sigma$ must be rank-2 tensors consisting of pure scalars at first-order, and so $\hat{\mathscr{V}}(\Dis\Del\Div\Div\T\sigma)$ and $\hat{\mathscr{T}}(\Dis\Del\Div\Div\T\sigma)$ must be gauge-invariant objects at second. In a similar fashion, one can form $\Div\Div\hat{\mathscr{V}}(\T\sigma)$ or $\Div\Div\hat{\mathscr{T}}(\T\sigma)$, which must vanish at first-order and so be gauge-invariant scalars at second-order.  In an accompanying paper this is extended into a general technique for defining gauge-invariant SVT  objects at any order~\cite{C}.

\appendix
\begin{widetext}

\section{The 1+3 covariant approach in index-free notation}

Index free notation may be used for any equations which are irreducibly split, and all objects appearing are similarly split~\cite{Marklund:2004qz}. In this case we denote a 3-vector $V^a$ by $\V{V}$ and, more generally, a PSTF tensor $X_{a\cdots b}$ by $\T{X}$. This notation is unambiguous provided the valance of an equation or variable is known.

We define three products between vectors and PSTF tensors. For this, let $\V{V}, \V{W}$ be rank-1, $\T{X}, \T{Y}$ be rank-2 and $\T{\Phi}, \T{\Psi}$ be rank-3 (rank-3 objects commonly appear as distortions of rank-2 tensors): 
\\
\small
\begin{center}
\begin{tabular}{|l|c|c|c|c|c|}\hline
Type  &Rank 0  &  Rank 1& Rank 2&Rank 3& Rank 4 \\ \hline
dot& $V^aW_a=\V{V}\cdot\V{W}$& $V^bX_{ab}=\V{V}\cdot\T{X}$  & $\Phi_{abc}V^{c}=\T{\Phi}\cdot\V{V}$& &  \\
product& $X^{ab}Y_{ab}=\T{X}\cdot\T{Y}$& $\Phi_{abc}X^{bc}=\T{\Phi}\cdot\T{X}$&  &  &\\
&$\Phi_{abc}\Psi^{abc}=\T{\Phi}\cdot\T{\Psi}$& & & &
\\\hline
cross &  &$\epsilon_{abc}V^bW^c=\V{V}\times\V{W}$& $\epsilon_{cd\<a}V^c{X^d}_{b\>}=\V{V}\times\T{X}$ & $\epsilon_{de\<a}V^d{\Phi^e}_{bc\>}=\V{V}\times\T{\Phi}$ & $\epsilon_{ef\<a}{X^{e}}_b{\Phi^f}_{cd\>}=\T{X}\times\T{\Phi}$ \\
product& &$\epsilon_{abc}{{X}^{b}}_{d}Y^{cd}=\T{X}\times\T{Y}$&  $\epsilon_{cd\<a}X^{ce}{\Phi^d}_{b\>e}=\T{X}\times\T{\Phi}$& $\epsilon_{de\<a}{X^{d}}_b{Y^e}_{c\>}=\T{X}\times\T{Y}$ &\\
& &$\epsilon_{abc}{\Phi^{b}}_{ef}\Psi^{cef}=\T{\Phi}\times\T{\Psi}$& & &\\\hline
circle & & &$V_{\langle a}W_{b\rangle}=\T{V}\circ\T{W}$&$X_{\<ab}V_{c\>}=\bm X\circ\bm V$& $X_{\<ab}Y_{cd\>}=\bm X\circ\bm Y$ \\
product& & &${X_{\langle a}}^{c}Y_{b\rangle c}=\T{X}\circ\T{Y}$&${\Phi_{\<ab}}^{d}X_{c\>d}=\bm \Phi\circ\bm X$& 
$\Phi_{\<abc}V_{d\>}=\bm \Phi\circ\bm V$\\
& & &$\Phi_{cd\langle a}{\Psi^{cd}}_{b\rangle}=\T{\Phi}\circ\T{\Psi}$&  &$\Phi_{e\langle ab}{\Psi^{e}}_{cd\rangle}=\T{\Phi}\circ\T{\Psi}$\\\hline
\end{tabular}\\[5mm]
\end{center}
\normalsize

\noindent Note that the notation is unambiguous once we know the rank of the object. For example, $\T{\Phi}\circ\T{\Psi}$ can mean a rank 2, 4 or 6 tensor. However, since this notation is only used on PSTF tensors, once the rank is specified, the product is unique. (Indeed, we don't actually need the new symbol $\circ$ as $\cdot$ would do just as well~-- we have introduced it for extra clarity.) Note also that the cross product is anti-symmetric on its arguments.

In the fully non-linear equations of the covariant formalism we normally use $\Del_a$ rather than $\sdel_a$, which is only really useful on an FLRW background. Let us also define:
\ba
\sdel&=&a\Del,\\
\label{relation}\curl &=&a\Curl,\\
\div &=&a\dv,\\
\dis &=& a\Dis\,.
\ea
With this notation, all the identities above (and below) which depend on spatial commutation relations, are the same, if we replace $K\mapsto-\frac{1}{6} {}^3\!R= K/a^2$ on an FLRW background. The only difference comes when commuting time derivatives. 

The 1+3 evolution and constraint equations are in this notation:

\noindent Evolution equations:

\noindent Rank 0:
\ba 3\dot{H}-\dv \V{A} &=& -3H^2 + \V{A} \cdot \V{A}
-{\T{\sigma}} \cdot {\T{\sigma}} +2 \V{\omega} \cdot \V{\omega}
-\:{1}{2}\left(\rho +3p \right) + \Lambda\\
\nonumber\\
\dot{\rho} + \dv \V{q} &=& -3H\left(\rho +p \right)
-2\V{A} \cdot \V{q} - {\T{\sigma}} \cdot {\T{\pi}}
\ea
\noindent Rank 1:
\ba
\label {rk11}\dot{\V{\omega}} -\:{1}{2}\Curl \V{A} &=& -2H\V{\omega}
+\V{\omega} \cdot {\T{\sigma}}\\
\nonumber\\
\dot{\V{q}} +\Del p + \dv {\T{\pi}} &=& -4H \V{q} -
\V{q} \cdot {\T{\sigma}} -\left(\rho+p\right)\V{A}
-\V{A}\cdot{\T{\pi}} -\V{\omega}\times \V{q}
\ea
\noindent Rank 2:
\ba
\dot{{\T{\sigma}}} -\Dis \V{A} &=& -2H{\T{\sigma}}
+\V{A} \circ \V{A} -{\T{\sigma}} \circ {\T{\sigma}} -
\V{\omega} \circ \V{\omega} -{\T{E}} +\:{1}{2}{\T{\pi}} \\
\nonumber\\
\label{rank22}\dot{{\T{E}}} +\:{1}{2}\dot{\T{\pi}} -\Curl {\T{H}}+\:{1}{2}\Dis \V{q} &=&
 -\:{1}{2}
\left(\rho+p\right){\T{\sigma}}
-3H\left({\T{E}} +\:{1}{6}{\T{\pi}} \right)\nonumber\\
&&~~~+3\,{\T{\sigma}} \circ {\T{E}}-\:{1}{2} {\T{\sigma}}
\circ {\T{\pi}} - \V{A} \circ \V{q} + 2 \V{A} \times {\T{H}}
+\V{\omega} \times {\T{E}} + \:{1}{2} \V{\omega}
\times {\T{\pi}}\\
\nonumber\\
\label{rank23}\dot{{\T{H}}} +\Curl {\T{E}}-\:{1}{2}\Curl{\T{\pi}} &=& 
-3H {\T{H}} + 3\,{\T{\sigma}}\circ {\T{H}}
+\:32 \V{\omega} \circ \V{q} - 2\, \V{A} \times {\T{E}}
-\:{1}{2}\V{q}\times{\T{\sigma}} +\V{\omega}\times {\T{H}}
\ea

\noindent{Constraint equations:}

\noindent Rank 0:
\ba
0 &=& \dv\V{\omega} - \V{A}\cdot\V{\omega}
\ea
\noindent Rank 1:
\ba
0 &=& \dv{\T{\sigma}} -2\Del {{H}}
+ \Curl\V{\omega}+2\,\V{A}\times\V{\omega}+\V{q}.\\
\nonumber\\
0 &=& \dv {\T{E}} +\:{1}{2}\dv{\T{\pi}}
-\:{1}{3}\Del\rho +H\V{q} -\:{1}{2}
{\T{\sigma}}\cdot \V{q} -3\,\V{\omega}\cdot {\T{H}}
-{\T{\sigma}}\times {\T{H}} +\:{3}{2}\V{\omega}
\times \V{q}\\
\nonumber\\
\label{conR1}0 &=& \dv {\T{H}} +\:{1}{2}\Curl \V{q} +\left(\rho+p\right)\V{\omega}
+3\,\V{\omega}\cdot{\T{E}}-\:{1}{2}\V{\omega}\cdot{\T{\pi}}
 + {\T{\sigma}}\times
{\T{E}}+\:{1}{2}{\T{\sigma}}\times{\T{\pi}}
\ea
\noindent Rank 2:
\ba
\label{con1}0&=& {\T{H}} +\Dis\V{\omega}
-\Curl{\T{\sigma}}+ 2\,\V{A}\circ\V{\omega}
\ea
In this notation it is actually much easier to manipulate the equations, and write down the various commutation relations. For example, 
\ba
(\dv\V{V}\dot)&=&\dv\dot{\V{V}}+\V A\cdot\dot{\V V} -H\dv\V V -\V\sigma\cdot\Dis\V V-\V\omega\cdot\Curl\V V+2H\V A\cdot\V V 
\nonumber\\&& -\V A\cdot\T\sigma\cdot\V V-\V A\cdot(\V\omega\times\V V) 
-\V V\cdot\dv\T\sigma+2\V V\cdot \Del H-\V V\cdot\Curl \V\omega\,,
\ea
is a fully non-linear commutation relation.


\subsection{Identities}

The following identities are useful for manipulating and expanding derivatives of products. They extend the usual rules of vector calculus to PSTF tensors. For example, the first three are normally given as vector calculus identities; here we have expanded them further using $\dis$. We use $\V{V},\V{W}$ for rank-1 tensors and $\T{X},\T{Y}$ for rank-2. We didn't quite have the energy to go further but products of rank-3 are actually required for full decomposition of source terms in rank-2 wave equations. Note that these are fully non-linear identities as derivatives are never commuted. This means that they apply equally well for $\Del$ as $\sdel$.

\ba
\sdel \left(\V{V}\cdot \V{W}\right)&=& \sdel_{a} \left(V^b
W_b\right) \nonumber\\
&=& \frac{1}{3}\left(\V{V} \div \V{W} + \V{W} \div \V{V} \right)
-\frac{1}{2}\left(\V{V} \times\curl \V{W} + \V{W} \times\curl \V{V}
\right) + \V{V}\cdot\dis \V{W}+ \V{W}\cdot\dis \V{V},
\ea
\ba
\div\left(\V{V}\times\V{W}\right)&=&\sdel^a\left(\epsilon_{abc}V^bW^c\right)\nonumber\\
&=& \V{W}\cdot\curl\V{V}-\V{V}\cdot\curl\V{W},
\\
\curl\left(\V{V}\times\V{W}\right)&=& \epsilon_{abc} \sdel^{b}\left(\epsilon^{cde}V_dW_e\right)\nonumber\\
&=& \frac{2}{3}\left(\V{V}\div\V{W}-\V{W}\div\V{V}\right)
+\frac{1}{2}\left(\V{V}\times\curl\V{W}-\V{W}\times\curl\V{V}\right)
-\left(\V{V}\cdot\dis\V{W}-\V{W}\cdot\dis\V{V}\right)
\\
\dis\left(\V{V}\times\V{W}\right)&=& \sdel_{\<a}\left(\epsilon_{b\>cd}V^cW^d\right)\nonumber\\
&=& \frac{1}{2}\left(\V{V}\circ\curl\V{W}-\V{W}\circ\curl\V{V}\right)
+\V{V}\times\dis\V{W}-\V{W}\times\dis\V{V}
\ea
\ba
\div \left(\V{V}\circ \V{W}\right)&=& \sdel^{a} \left(V_{\langle a}
W_{b\rangle}\right) \nonumber\\
&=& \frac{5}{9}\left(\V{V} \div \V{W} + \V{W} \div \V{V} \right)
-\frac{1}{12}\left(\V{V}
\times\curl \V{W} + \V{W} \times\curl \V{V} \right) \nonumber\\ &&~~~ 
+\frac{1}{6}\left( \V{V}\cdot\dis \V{W}+ \V{W}\cdot\dis
\V{V}\right)
%
\\\label{B19}
\curl \left(\V{V}\circ \V{W}\right)&=& 
\frac{1}{2}\epsilon_{cd\<a}\sdel^c \left(V_{b\>}W^d+V^dW_{b\>}\right) 
\nonumber\\
&=& \frac{3}{4}\left(\V{V}\circ\curl\V{W}+\V{W}\circ\curl\V{V}\right)
-\frac{1}{2}\left(\V{V}\times\dis\V{W}+\V{W}\times\dis\V{V}\right)
\\\label{B20}
\dis \left(\V{V}\circ \V{W}\right)&=&\sdel_{\<a}\left(V_{\<b}W_{c\>\>}\right)\nonumber\\
&=&\V{V}\circ\dis\V{W}+\V{W}\circ\dis\V{V} 
\ea
\ba
\label{B21}
\div \left(\V{V}\cdot {\T{X}}\right)&=& \sdel^{a} \left(V^b
X_{ab}\right) \nonumber\\
&=& {\T{X}}\cdot\dis \V{V} + \V{V}\cdot\div {\T{X}}.
\\\label{B22}
\curl \left(\V{V}\cdot {\T{X}}\right)&=& \epsilon_{abc} \sdel^{b}
\left(V^d
X^{c}{}_{d}\right) \nonumber\\
&=& -\frac{1}{2}{\T{X}}\cdot\curl \V{V}  + \V{V} \cdot \curl
{\T{X}} -  {\T{X}}\times \dis \V{V}
\\\label{B23}
\dis \left(\V{V}\cdot {\T{X}}\right)&=& \sdel_{\langle a} V^c
X_{b\rangle c} \nonumber\\
&=& \frac{1}{3}{\T{X}}\div \V{V} 
- \frac{1}{2}
{\T{X}}\times\curl \V{V}  
+ {\T{X}} \cdot \dis \V{V} + \frac{3}{10} \V{V} \circ
\div {\T{X}} - \frac{1}{3} \V{V} \times \curl {\T{X}} + \V{V}
\cdot \dis {\T{X}},
\ea
\ba
\label{B24}
\div\left(\V{V}\times\T{X}\right)&=&\sdel^a\left(\epsilon_{cd\<a}V^cX^d_{~b\>}\right)\nonumber\\
&=& \frac{3}{4}\T{X}\cdot\curl\V{V}-\frac{1}{2}\V{V}\cdot\curl\T{X}-\frac{1}{2}\T{X}\times\dis\V{V}+\frac{1}{2}\V{V}\times\div\T{X},
\\
\label{B25}
\curl\left(\V{V}\times\T{X}\right)&=& 
	\frac{1}{2}\epsilon_{cd\<a}\sdel^c\left(\epsilon_{|ef|~}^{~~~~d}V^eX^f_{~b\>}+\epsilon_{|ef|b\>}V^eX^{df}\right)\nonumber\\
	&=&-\frac{1}{2}\T{X}\div\V{V}+\frac{1}{4}\T{X}\curl\V{V}-\frac{1}{3}\V{V}\curl\T{X}+\frac{1}{2}\T{X}\circ\dis\V{V}-\frac{1}{2}\V{V}\cdot\dis\T{X},
\\
\label{B26}
\dis\left(\V{V}\times\T{X}\right)&=&\sdel_{\<c}\left(\epsilon_{|de|\<a}V^dX^e_{~b\>\>}\right)\nonumber\\
&=&\frac{1}{3}\V{V}\circ\curl\T{X}+\T{X}\circ\curl\V{V}+\V{V}\times\dis\T{X}-\T{X}\times\dis\V{V} ,
\ea
\ba
\label{B27}
\sdel \left({\T{X}}\cdot {\T{Y}}\right)&=& \sdel_{c}
\left(X^{ab}
Y_{ab}\right) \nonumber\\
&=& \frac{3}{5}\left(\div {\T{X}}\cdot {\T{Y}} + \div
{\T{Y}}\cdot {\T{X}} \right) +\frac{2}{3}\left({\T{X}}
\times\curl {\T{Y}} + {\T{Y}} \times\curl {\T{X}} \right) +
{\T{X}}\cdot\dis {\T{Y}}+ {\T{Y}}\cdot\dis {\T{X}},
\ea
\ba
\label{B28}
\div\left(\T{X}\times\T{Y}\right)&=&\sdel^a\left(\epsilon_{abc}X^b_{~d}Y^{cd}\right)\nonumber\\
&=& -\T{X}\cdot\curl\T{Y}+\T{Y}\cdot\curl\T{X},
\\
\label{curlxxy}
\curl\left(\T{X}\times\T{Y}\right)&=&\epsilon_{abc}\sdel^b\left(\epsilon^{cde}X_{df}Y_e^{~f}\right)\nonumber\\
&=&-\frac{7}{10}\left(\T{X}\cdot\div\T{Y}-\T{Y}\cdot\div\T{X}\right)
+\frac{1}{3}\left(\T{X}\times\curl\T{Y}-\T{Y}\times\curl\T{X}\right)
-\T{X}\cdot\dis\T{Y}+\T{Y}\cdot\dis\T{X},
\\\label{B29}
\dis\left(\T{X}\times\T{Y}\right)&=&\sdel_{\<a}\left(\epsilon_{b\>cd}X^{cf}Y^d_{~f}\right)\nonumber\\
&=&\frac{3}{10}\left(\T{X}\times\div\T{Y}-\T{Y}\times\div\T{X}\right)
+\frac{1}{3}\left(\T{X}\circ\curl\T{Y}-\T{Y}\circ\curl\T{X}\right)
-\T{X}\times\dis\T{Y}+\T{Y}\times\dis\T{X},
\ea
\ba
\label{B30}\div\left(\T{X}\circ\T{Y}\right)&=&\sdel^a\left(X_{\<a}^{~~d}Y_{b\>d}\right)\nonumber\\
&=&\frac{9}{20}\left(\T{X}\cdot\div\T{Y}+\T{Y}\cdot\div\T{X}\right)
-\frac{5}{18}\left(\T{X}\times\curl\T{Y}+\T{Y}\times\curl\T{X}\right)
+\frac{1}{6}\left(\T{X}\cdot\dis\T{Y}+\T{Y}\cdot\dis\T{X}\right),
\\
\label{B31}\curl\left(\T{X}\circ\T{Y}\right)&=&\epsilon_{cd\<a}\sdel^c\left(\frac{1}{2}X^{de}Y_{b\>e}+\frac{1}{2}X_{b\>}^{~~e}Y^d_{~~e}\right)\nonumber\\
&=&-\frac{3}{20}\left(\T{X}\times\div\T{Y}+\T{Y}\times\div\T{X}\right)
+\frac{1}{6}\left(\T{X}\circ\curl\T{Y}+\T{Y}\circ\curl\T{X}\right)
-\frac{1}{2}\left(\T{X}\times\dis\T{Y}+\T{Y}\times\dis\T{X}\right),
\\
\label{B32}\dis\left(\T{X}\circ\T{Y}\right)&=&\sdel_{\<a}\left(X^d_{~\<b}Y_{c\>\>d}\right)\nonumber\\
&=&\frac{3}{10}\left(\T{X}\circ\div\T{Y}+\T{Y}\circ\div\T{X}\right)
+\frac{1}{3}\left(\T{X}\times\curl\T{Y}+\T{Y}\times\curl\T{X}\right)
\T{X}\circ\dis\T{Y}+\T{Y}\circ\dis\T{X}
\ea

\end{widetext}

\acknowledgments

We thank Kishore Ananda who participated in early stages of this work,  David Wands for discussions, and Bishop Mongwane and David Wiltshire for comments. This work is funded by the NRF (South Africa).

\end{document}